\input{aipcheck}

\documentclass[
    ,final            % use final for the camera ready runs
  ]
  {aipproc}

\layoutstyle{6x9}

\begin{document}

\title{A simple determination of some characteristics of  the $\beta$ Pictoris system }

\classification{PACS:95.30.Qd , 97.82.Jw , 97.82-j}
                %\texttt{http://www.aip..org/pacs/index.html}>}
\keywords      {Plasmas astrophysical, debris disks, protoplanetary disks, beta Pictoris}

\author{Vladan \v Celebonovi\'c}{
  address={{Inst.of Physics,Pregrevica 118,11080 Zemun-Beograd,Serbia }\\
$vladan@phy.bg.ac.yu$}}

\begin{abstract}
The aim of this contribution is to determine the lower limit  of the electron number density, the Debye temperature and the specific heat of the solid grains present in the protoplanetary cloud around the star $\beta$ Pictoris. The calculation has been performed using the Salpeter criterion,slightly modified a couple of years ago. The results obtained are physically reasonable,and could be helpful in determining the chemical composition of the solid particles in this protoplanetary system.
% This template file shows how to use the \texttt{aipproc} class to
% produce a paper with the correct layout for \emph{%
%   AIP Conference Proceedings  6in   x 9in single column}.

% A full description of the features supported by the \texttt{aipproc}
% class can be found in the \texttt{aipguide.pdf} document accompanying
% the distribution.

% Frequently asked questions can be found in the \texttt{FAQ.txt}
% document.
\end{abstract}

\maketitle

%%%%%%%%%%%%%%%%%%%%%%%%%%%%%%%%%%%%%%%%%%%%
%% MAINMATTER
%%%%%%%%%%%%%%%%%%%%%%%%%%%%%%%%%%%%%%%%%%%%

\section{Introduction}
The existence of a planetary system in which we live is known to mankind at least since antiquity. First attempts aiming at finding a rational scientific explanation of the origin of our planetary system date back to the $XVIII$ century, and they are due to work  by P.S.Laplace and E.Kant. Although termendous progress has been made, this problem is far from being clarified in all the details. For recent reviews of the subject,see for example \cite{BRUSH:90} or \cite{PAPA:06}.

 In the "pre-satellite" period observational confirmations of the existence of protoplanetary clouds were hard to obtain, mainly because they are theoretially expected to radiate weakly in the infrared and therefore almost unobservable from ground based observatories. Real confirmation had to await the launch of infrared satelltes such as $IRAS$ in the eighties of the last century. The archive of data from $IRAS
 $ can be accessed at the following web address:  \url{http://irsa.ipac.caltech.edu/IRASdocs/iras.html}. 
 
 The first protoplanetary disk to be discovered using $IRAS$ was the disk around the bright star $\alpha$ Lyrae \cite{AUMM:84},and immediately after that the one around $\beta$ Pictoris \cite{SMTE:84}. How important these two discoveries are is illustrated by the number of papers which followed after them.As a sequel to \cite{SMTE:84} nearly $10000$ publications have appeared. 
 
 The aim of this contribution is to present a simple calculation of the lower limit of the electron number density in the cloud around $\beta$ Pictoris and of the Debye temperature of the solid grains in it. As the Debye temperature is material dependent,this value combined with spectroscopic data,could be helpful in the determination of the chemical composition of the grains. The calculation to be reported in the next section is based upon a theory proposed in \cite{VLDA:02}.

\section{The calculations}

\subsection{Theoretical basis}
Phase transitions (PT)  of various kinds are active contributors to changes in the world around us. One particular kind of a PT is the so called plasma-solid (PS) phase transition. Physical reasoning shows that a PS transition certainly occurs in astrophysical settings such as the proto-planetary and proto-stellar clouds. 

It is  known that the universe contains various kinds of plasmas.On the other hand,it is also known that solid objects are present in the universe,so there must exist some region in  phase space where a transition between the two regimes takes place. 

A useful theoretical tool for the study of the PS transition is a criterion proposed by Salpeter \cite{SALP:61}. He made a thorough analysis of a zero-temperature plasma. In a part of his paper he considered a system of positive ions of given charge and mass rigidly fixed in the nodes of a perfect crystal lattice,and {\it estimated} their zero point energy. The behaviour of such a system can be described by the ratio 
\begin{equation}
	f_{S}=\frac{E_{z,p}}{E_{C}}
\end{equation}
where $E_{z,p}$ denotes the zero point energy of the ions and $E_{C}$ is the Coulomb energy. Salpeter showed that a PS transition occurs for $f=1$. In \cite{VLDA:02} this criterion was somewhat modified,in the sense that the mean enegies per particle of a solid $E_{p,s}$ and of a Fermi gas $E_{p,e}$ were compared,so that 
\begin{equation}
	f_{m}=\frac{E_{p,s}}{E_{p,e}}
\end{equation}
As in the original version,the PS transition occurs for $f_{m}=1$. A general expression for this ratio was obtained in \cite{VLDA:02}. It was shown that up to and including second order terms in $(\theta/T)$ where $\theta$ is the Debye temperature, the ratio $f_{m}$ is given by
\begin{equation}
	f_{m}\cong \frac{9n}{2KT} \frac{20T^{2}+\theta^{2}}{12A^{2}n^{4/3}+5\pi^{2}T^{2}} 
\end{equation}
where $K=gm^{3/2}/2^{1/2}\pi^{2}\hbar^{3}$, $g=2s+1$, $m$ is the particle mass and $A=(3\pi^{2})^{2/3}\hbar^{2}/2m$. Inserting $f_{m}=1$ into this expression one can solve it for $\theta$.Developing the result thus obtained into series in $T$ (which is plausible because of the small values of temperatures involved) and retaining only the biggest terms,one finally gets that
\begin{equation}
	\theta \cong67388 (\frac{T^{5}}{n^{3}})^{1/2}
\end{equation}
It can be shown  \cite{VLDA:02} that this result is valid for 
\begin{equation}
	T\leq0.0021 n
\end{equation}
  
\subsection{Application to $\beta$ Pictoris} 
$\beta$ Pictoris has been observed by the $FUSE$ satellite in $2001$ and $2002$ \cite{NATU:06}. Analysis of the absorption doublet $CII$ $\lambda1037$ showed that the temperature in the circumstellar cloud is $T\cong132$ $K$. 

Inserting the observed value of the temperature into eq.(5) it follows that $n\geq 6.3\times10^{4}$. Eq.(4) then gives that $\theta\leq850$ $K$ which is the upper limit to the value of the Debye temperature of the material making up the solid grains in the cloud around $\beta$ pictoris. 

Consulting tabulated data on the Debye temperature of materials opens the way to interesting conclusions. Namely,the value of $\theta$ for carbon is $2230$ $K$ while the correspondning value for $Si$ is only $645$ $K$. This means that a compound containing carbon and silicon should have a Debye temperature less than $850$ $K$. The chemical formula of this compound would be of the form $C_{m}Si_{n}$,where $m$ and $n$ would have to be determined by a separate chemical analysis.

Pushing the reasoning further, once the value of the Debye temperature is known, the specific heat of a material can be calculated as: 
\begin{equation}
	C_{V} = \frac{12\pi^{4}}{5} k_{B} N_{A}(\frac{T}{\theta})^{3}
\end{equation}
Inserting eq.(4) into eq.(6) leads to 
\begin{equation}
	C_{V}\cong (\frac{n}{T})^{9/2}
\end{equation}
This means that the specific heat of the solid grains in a plasma in which a PS is possible is proportional to the ratio of the plasma density and temperature. Now,both $n$ and $T$ are some functions of the radial distance from the central star,but the problem is that they are extremely difficult to determine from observation. They can be approximately estimated from theoretical models of protoplanetary clouds,such as the recent model proposed in \cite{GALI:06}. 

Photometry can provide the radial distribution of brightness of a protoplanetary cloud,but again a theoretical model of a cloud is needed to "transit" from the brightness distribution to the distribution of temperature or number density.

\subsection{Conclusions} 
In this contribution we have analyzed in some detail the protoplanetary disk around the star $\beta$ Pictoris. This disk was discovered by the infrared satellite IRAS. It was analyzed using the modernized form of the criterion for the occurence of a plasma-solid transition proposed by Salpeter. 

Starting from the observed value of the temperature,the lower limit of the number density and the upper limit of the Debye temperature of the solid grains within the cloud were determined. Hints on the chemical composition of the grains have been made. The analytical form of the specific heat of the material making up the solid grains was also determined. Details of this expression will be discussed elsewhere. Knowledge of the specific heat of the solid grains is important for planetary formation studies,especially for studies of transport of heat in a protoplanetary disk.

%%%%%%%%%%%%%%%%%%%%%%%%%%%%%%%%%%%%%%%%%%%%
%% Sample figure:
%%
%% The option [height=...] scales the picture to the given height,
%% without it it would be printed at its nominal size
%%%%%%%%%%%%%%%%%%%%%%%%%%%%%%%%%%%%%%%%%%%%

%begin{figure}
% \includegraphics[height=.3\textheight]{golfer}
% \caption{Picture to fixed height}
%end{figure}

%%%%%%%%%%%%%%%%%%%%%%%%%%%%%%%%%%%%%%%%%%%%
%% SAMPLE TABLE
%%
%% Shows the use of \tablehead and \tablenote
%% macros
%%%%%%%%%%%%%%%%%%%%%%%%%%%%%%%%%%%%%%%%%%%%

%\begin{table}
%\begin{tabular}{lrrrr}
%\hline
%  & \tablehead{1}{r}{b}{Single\\outlet}
%  & \tablehead{1}{r}{b}{Small\tablenote{2-9 retail outlets}\\multiple}
%  & \tablehead{1}{r}{b}{Large\\multiple}
%  & \tablehead{1}{r}{b}{Total}   \\
%\hline
%1982 & 98 & 129 & 620    & 847\\
%1987 & 138 & 176 & 1000  & 1314\\
%1991 & 173 & 248 & 1230  & 1651\\
%1998\tablenote{predicted} & 200 & 300 & 1500  & 2000\\
%\hline
%\end{tabular}
%\caption{Average turnover per shop: by type
%  of retail organisation}
%\label{tab:a}
%\end{table}

%\begin{enumerate}
%\item
%  An item \cite{Liang:1983}
%\item
%  Another item with sub entries
%  \begin{enumerate}
%  \item
%   A sub entry \cite{Wang}
%  \item
%   Second sub entry
%  \end{enumerate}
%\item
%  The final item with normal label.
%\end{enumerate}

%\begin{description}
%\item[Infandum]
 
%\item[Sed]

%\item[Lamentabile] regnum cruerint Danai; quaeque ipse miserrima vidi, et

%\end{description}

%\section{<A section>}

%%%%%%%%%%%%%%%%%%%%%%%%%%%%%%%%%%%%%%%%%%%%%%%%
%% BACKMATTER
%%%%%%%%%%%%%%%%%%%%%%%%%%%%%%%%%%%%%%%%%%%%%%%%

\begin{theacknowledgments}
 This contribution was prepared within the research project $141007$ financed by the Ministry of Science and Protection of Environement of Serbia.  
\end{theacknowledgments}


.

\end{document}